\documentclass[fleqn,10pt]{wlscirep}
\usepackage[utf8]{inputenc}
\usepackage[T1]{}

\title{Image complexity based fMRI-BOLD visual network categorization across visual datasets using topological descriptors and deep-hybrid learning}

\author[1,*]{Debanjali Bhattacharya}
\author[1]{Neelam Sinha}
\author[1]{Yashwanth R}
\author[1]{Amit Chattopadhyay}
\affil[1]{International Institute of Information Technology (IIIT), Bangalore, 560100, India}
\affil[*]{debanjali.bhattacharya@iiitb.ac.in, jhimli20@gmail.com}

\keywords{fMRI time-series, Correlation, Partial correlation, Visual network, Topological data analysis, Deep hybrid learning, Classification}
\newcommand{\R}{\mathbb{R}}
\newcommand{\Dg}{\mathit{Dg}}
\newcommand{\ExDg}{\mathit{ExDg}}
\newcommand{\figref}[1]{Figure \ref{#1}}

\begin{abstract}

This study proposes a new approach that investigates differences in topological characteristics of visual networks, which are constructed using fMRI BOLD time-series corresponding to visual datasets of COCO, ImageNet, and SUN. A publicly available BOLD5000 dataset is utilized that contains fMRI scans while viewing 5254 images of diverse complexities. The objective of this study is to examine how network topology differs in response to distinct visual stimuli from these visual datasets. To achieve this, 0- and 1-dimensional persistence diagrams are computed for each visual network representing COCO, ImageNet, and SUN. For extracting suitable features from topological persistence diagrams, K-means clustering is executed. The extracted K-means cluster features are fed to a novel deep-hybrid model that yields accuracy in the range of $90\%-95\%$ in classifying these visual networks. To understand vision, this type of visual network categorization across visual datasets is important as it captures differences in BOLD signals while perceiving images with different contexts and complexities. 
Furthermore, distinctive topological patterns of visual network associated with each dataset, as revealed from this study, could potentially lead to the development of future neuroimaging biomarkers for diagnosing visual processing disorders like visual agnosia or prosopagnosia, and tracking changes in visual cognition over time.

\end{abstract}

\begin{document}

\flushbottom
\maketitle

\thispagestyle{empty}

\section*{Introduction}
The human brain forms a complex neural circuit with approximately 86 billion neurons and 150 trillion synapses, that facilitates information pathways across different brain regions \cite{ref1}. Advancements in studying the intricate nature of the human brain have surged in recent decades. As a result, neuroscientists have emphasized the exploration of brain neuronal connectivity patterns, which provide vital insights into the structural connectivity (anatomical links), functional connectivity (statistical dependency), and effective connectivity (causal interactions) of the brain. Consequently, brain connectivity analysis has emerged as a focal point, captivating the attention of numerous researchers in the field of neuroscience \cite{ref2}. For this, functional Magnetic Resonance Imaging (fMRI) is widely used that exploits the changes in magnetic properties of oxygenated and de-oxygenated blood to measure the fluctuation in neuronal activities during well-defined tasks or resting-state conditions. \par
During over last two decades, several fMRI studies have been attempted to analyze the brain connectivity pattern in patients with neurological and psychiatric disorders as well as in healthy individuals, during different cognitive loads as well as during resting-state \cite{ref3, ref4,ref5,ref6,ref7,ref8,ref9,ref10,ref11,ref12,ref13}.  
However, studies on human visual perception have been still limited due to complex experimental procedures involved in generating an adequate number of good quality fMRI neuro-image data, representing vision. The release of the publicly available dataset “BOLD5000” \cite{ref14}, which contains fMRI scans of four different subjects acquired while viewing 5254 images, has made it possible to study brain dynamics during visual tasks in greater detail. These images were chosen from three well-known computer vision datasets: (i) COCO, (ii) ImageNet, and (iii) SUN, which contain images with diverse contexts and complexities and are widely used for bench-marking techniques in computer vision. 
The fMRI blood-oxygen-level-dependent (BOLD) time series signals have emerged as a powerful tool for detecting changes in brain activation and identifying dysfunctions. Additionally, one potential application of the BOLD time series is recognizing differences in visual perception. 
The central hypothesis driving this study is that \textit{BOLD signals in human brain exhibit variations when processing images with diverse complexities, object scales, and contextual information.} Thus, the main objective of the study is to address the question: \textit{can fMRI BOLD time series, obtained from active voxel locations in the brain, be effectively employed to categorize visual stimuli into their corresponding visual datasets?} 
Accomplishing this task would validate the notion that the brain processes visual information in a way that aligns with human-intuitive categorization. In other words, it could be interpreted as the response evoked while viewing images of varying degrees of complexity are also correspondingly distinct. 
The present study is the extension of our previous research works \cite{kancharala2023spatial,naveen}, which demonstrated the ability of deep neural networks to differentiate visual stimuli and classify them into COCO, ImageNet, and SUN datasets using both one-dimensional BOLD time-series signals and their 2D representations. 

\textit{Research gap and motivation for the current study:} Our previously reported work \cite{kancharala2023spatial,naveen} used long short-term memory (LSTM), Bi-directional LSTM and parallel convolutional neural network (CNN) architectures for classification. By visualizing LSTM and CNN features using t-distributed stochastic neighbor embedding (t-SNE), we showed clear differentiation of temporal and spatial features between BOLD time series of COCO, ImageNet, and SUN, facilitating the successful classification of BOLD time series into corresponding datasets. However, the study did not explore the ensuing network response to these stimuli which left the unaddressed question of whether stimuli having different contexts and complexities result in inducing distinct network connectivity patterns. This question is addressed in the current study where we report the investigation of the differences in the topological characteristics of the visual network, which represents COCO, ImageNet, and SUN. The main contributions of this study are as follows: 
\begin{enumerate}
    \item  Construction of a more reliable visual network by integrating both marginal correlation and partial correlation estimates of fMRI BOLD time series in order to capture direct associations between network nodes. 
    \item  Extraction of topological descriptors using persistence diagram and K-means clustering 
 to quantify changes across different visual networks, for varying scales or levels of connectivity.
    %aiming to examine the differences in topological characteristics of visual networks
    
    \item A new deep-hybrid model is proposed to effectively classify different visual networks across the three considered computer vision datasets using the extracted topological features, aiming to reveal how differently human brain functions while looking into images having different scales, contexts, and complexities. 
\end{enumerate}
By doing so, we aim to understand how the network topology differs in response to distinct visual stimuli from COCO, ImageNet, and SUN datasets.
To the best of the authors' knowledge, this is the first study that reports the application of topological data analysis (TDA) on fMRI BOLD time series to advance understanding of visual networks.

\section*{Results} 
\subsection*{Dataset Description}
The publicly available BOLD5000 dataset is used for this study \cite{ref14}. Four right-handed healthy volunteers (M:F 1:3)  were selected from Carnegie Mellon University who were graduate students in the age range of 24 to 27 years. fMRI scans are acquired while viewing 5,254 images of diverse categories. These images were selected from three classical computer vision databases as described below.
\begin{enumerate}[label=(\roman*)]
    \item \textit{Common objects in context (COCO)}: This is a standard benchmark dataset for multiple object detection and segmentation tasks of complex indoor and outdoor scenes. 2,000 images from the COCO dataset were selected in the study. These multiple objects in the COCO dataset depict interaction with other objects in a realistic context (for example, images depicting basic human social interactions).
    
    \item  \textit{Scene Understanding (SUN)}: This dataset contains real-world scene images of indoor and outdoor environments and is traditionally used for scene classification. A set of 1,000 scene images covering 250 categories were selected which inclined to be more panoramic, having no focus on specific objects. 

    \item \textit{ImageNet}: This dataset is used for the classification and localization of singular objects that are centered in real-world scenes. 1,916 images were selected from the ImageNet database that mainly focuses on a single object in the picture. Unlike COCO and SUN images, the single object in the ImageNet images is mostly placed at the center in order to distinguish it clearly from the image background. 
\end{enumerate}

\begin{figure}[ht]
\centering
\includegraphics[width=0.8\linewidth]{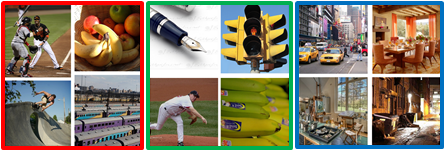}
\caption{Sample images, taken from the three computer vision datasets having different contents: \textit{Left:} COCO (\textit{Red box})- contains multiple objects and actions, \textit{Middle:} ImageNet (\textit{Green box})- contains single-focused object and \textit{Right:} SUN (\textit{Blue box})- contains indoor and outdoor scenes.}
\label{fig:imgs}
\end{figure}

The sample images from COCO, ImageNet, and SUN are shown in Figure~\ref{fig:imgs}. Out of $4,916$ images ($1,000$ from SUN + $2,000$ from COCO + $1916$ from ImageNet), $112$ images were randomly selected and repeated multiple times in the experiment. This resulted in a total of $5,254$ image presentations ($4,803$ unique images +$112$ images with $4$ times repetitions + $1$ image with $3$ times repetitions) for each participant which were chosen randomly during each trial of fMRI sessions and used as the experimental stimuli. 
Out of four subjects, three subjects completed $15$ functional sessions with all the $5,254$ image data, and the fourth subject could complete $9$ functional sessions with $3,108$ image data, due to discomfort in the fMRI. Each functional session consisted of $9$ to $10$ runs where in each run $37$ stimuli (images) were presented randomly to the participants. Each of the functional scanning sessions was roughly $1.5$ hours long for all subjects.
3T Siemens Verio MR scanner having a $32$-channel phased array head coil was used to acquire the fMRI data. The images were collected using a T2*W gradient recalled echo-planar imaging multi-band pulse sequence. To address the impact of signal overlap, voxel-specific GLM beta estimates are subjected to regularization using ridge regression. Further details on subject demographics, stimuli selection, fMRI scan acquisition, and data pre-processing procedures can be found in literature \cite{ref14}. 

In the present study, we perform two distinct approaches for the classification of visual networks across visual datasets - COCO, ImageNet, and SUN.
(1) Topological persistence diagram followed by K-means clustering (in the feature space of topological descriptors) to derive relevant features for quantifying the dissimilarities in the topological properties of visual networks of COCO, ImageNet, and SUN, constructed for each of the fMRI sessions.
(2) Subsequently, deep hybrid learning is employed to obtain compact embedding of K-means clustered features for better classification performance. \par

\subsection*{Improved classification using latent space features}
The initial approach for the classification of visual networks of COCO, ImageNet, and SUN, utilizes the XGBoost classifier which was given input of K-means clustering features computed from each of the persistence diagrams. XGBoost (Extreme Gradient Boosting) is a powerful ensemble machine learning algorithm, that combines the predictions of multiple base learners to produce a stronger prediction. XGBoost uses distributed, scalable gradient-boosted decision trees that provide parallel tree boosting, and instead of providing a single decision at each leaf node of the decision tree, it includes real-value scores of whether an instance belongs to a specific class. The decision is made by converting the scores into categories using a certain threshold only after the tree reaches its maximum depth. However, with 10-fold cross-validation, XGBoost classification on K-means features yields average accuracies of $50.11\%$, $69\%$, and $63.81\%$ for classifying COCO Vs. ImageNet, COCO Vs. SUN, and ImageNet Vs. SUN, respectively, across all subjects. Hence, to improve the classification performance, we have explored support vector machine (SVM) classification on autoencoder 4D latent space features, which led to significantly enhanced accuracies of $74.13\%$, $75.39\%$, and $70.4\%$, respectively. It is observed that the overall accuracy is increased in the range of $6.39\%$ (min.) to $23.72\%$ (max.) when utilizing the higher-order polynomial kernel.

\subsection*{Exploring PCA-reduced latent space embeddings}
To further enhance the classification performance, we applied Principal Component Analysis (PCA) on the autoencoder latent space features. Due to a limited number of observations, leave-one-out SVM classification is performed on the resulting PCA-reduced features which include 2D, 3D, and 4D representations, for various combinations of explained variances. The classification results across all subjects for different feature combinations are presented in Table~\ref{tab:table1}.
As shown in the table, incorporating PCA-reduced autoencoder latent space features led to a \emph{remarkable increase} in classification accuracy, ranging from $83\%$ to $95\%$ using the SVM-RBF kernel. Particularly, the best result is achieved when utilizing all PCA components, where we obtained maximum accuracies of $95\%$, $90\%$, and $91\%$ for classifying COCO Vs. ImageNet, COCO Vs. SUN, and ImageNet Vs. SUN, respectively. This signifies the effectiveness of PCA in improving the discriminative power of the autoencoder latent space features and yielding impressive classification performance.

\subsection*{Comparison with previous baseline studies}
As mentioned earlier, the current study is an extension of our previous baseline studies \cite{kancharala2023spatial,naveen} where we directly employed the raw BOLD time series corresponding to COCO, ImageNet, and SUN datasets for classification. In contrast, the current work utilizes visual networks, constructed using BOLD time series and examines their distinctive topological characteristics for improved classification.
Constructing a network based on correlations is computationally more efficient than analyzing raw time series data, especially for large datasets like BOLD5000. The connectivity patterns captured by the graph are more generalizable across individuals, leading to more reliable estimates. Hence, the current approach provides more meaningful insights into the underlying connectivity patterns and offers promising avenues for future research in brain network analysis.
While comparing the classification results with our baseline studies, we observe a $4\%$ and $7\%$ increase in accuracy when classifying COCO vs. ImageNet uses 3D and 4D PCA-reduced latent space features, respectively. 
However, the performance in classifying COCO vs. SUN and ImageNet vs. SUN decreased to $90\%$ when employing PCA-derived autoencoder latent-space features instead of the direct BOLD time series signal. In the current study, despite the high variability observed across subjects, across sessions, and the reduction in the number of instances due to the graph-based approach, we believe that the topological analysis of visual networks yields superior performance in distinguishing visual networks for COCO, ImageNet, and SUN datasets.
%The high accuracy of 99\% while using direct BOLD time series could be due to their specific sensitivity to sequential patterns and temporal dependencies. 

\begin{table}[ht]
\centering
\begin{tabular}{|l|l|l|l|l|l|l|}
\hline
Types of features & ML  &Visual &Precision &Recall &F1-score &Accuracy \\
(dimensions) & Classifier & Network & & & & \\
\hline
K-means (12D) &XGBoost & COCO &0.48$\pm$0.01 &0.54$\pm$0.03	&0.51$\pm$0.01	&0.50$\pm$0.02 \\
& & ImageNet &0.51$\pm$0.02 &0.47$\pm$0.05 &0.50$\pm$0.07 & \\
\cline{3-7}
 & & COCO &0.67$\pm$0.08 &0.72$\pm$0.04	&0.70$\pm$0.03	&0.69$\pm$0.06 \\
& & SUN &0.74$\pm$0.05 &0.66$\pm$0.09	&0.68$\pm$0.08 & \\
\cline{3-7}
& & ImageNet &0.62$\pm$0.06 &0.67$\pm$0.09	&0.64$\pm$0.08	&0.63$\pm$0.07 \\
& & SUN &0.66$\pm$0.09 &0.60$\pm$0.1	&0.62$\pm$0.05 & \\
\hline
\hline

Autoencoder &SVM  & COCO &0.72$\pm$0.04 &0.77$\pm$0.02	&0.75$\pm$0.01	&0.74$\pm$0.02 \\
latent space (4D) &(Polynomial) & ImageNet &0.75$\pm$0.01 &0.70$\pm$0.06 &0.73$\pm$0.04 & \\
\cline{3-7}
& & COCO &0.74$\pm$0.08 &0.78$\pm$0.06	&0.76$\pm$0.06	&0.75$\pm$0.07 \\
& & SUN &0.77$\pm$0.07 &0.72$\pm$0.1	&0.74$\pm$0.08 & \\
\cline{3-7}
& & ImageNet &0.69$\pm$0.07 &0.74$\pm$0.1	&0.71$\pm$0.07	&0.70$\pm$0.07 \\
& & SUN &0.73$\pm$0.09 &0.67$\pm$0.1	&0.69$\pm$0.07 & \\
\hline
\hline

Autoencoder+ & SVM& COCO &0.85$\pm$0.09 &0.87$\pm$0.05	&0.86$\pm$0.05	&0.86$\pm$0.05 \\
PCA (2D) & (rbf) & ImageNet &0.87$\pm$0.04 &0.84$\pm$0.1	&0.85$\pm$0.06 & \\
\cline{3-7}
& & COCO &0.85$\pm$0.06 &0.81$\pm$0.05	&0.83$\pm$0.04	&0.84$\pm$0.04 \\
& & SUN &0.82$\pm$0.04 &0.85$\pm$0.06	&0.84$\pm$0.04 & \\
\cline{3-7}
& & ImageNet &0.84$\pm$0.12 &0.82$\pm$0.12	&0.83$\pm$0.08	&0.83$\pm$0.09 \\
& & SUN &0.83$\pm$0.12 &0.84$\pm$0.12	&0.83$\pm$0.09 & \\
\hline
Autoencoder+ &SVM & COCO &0.91$\pm$0.08 &0.93$\pm$0.05	&0.92$\pm$0.06	&0.92$\pm$0.06 \\
PCA (3D) &(rbf) &ImageNet &0.93$\pm$0.05 &0.90$\pm$0.09	&0.91$\pm$0.06 & \\
\cline{3-7}
& & COCO &0.90$\pm$0.08 &0.85$\pm$0.06	&0.87$\pm$0.05	&0.88$\pm$0.05 \\
& & SUN &0.86$\pm$0.05 &0.90$\pm$0.08	&0.88$\pm$0.05 & \\
\cline{3-7}
& & ImageNet &0.92$\pm$0.1 &0.83$\pm$0.08	&0.87$\pm$0.09	&0.88$\pm$0.08 \\
& &  SUN &0.84$\pm$0.07 &0.92$\pm$0.09	&0.88$\pm$0.08 & \\
\hline
Autoencoder+ &SVM & COCO &0.93$\pm$0.05 &0.97$\pm$0.05	&0.95$\pm$0.04	&\textbf{0.95$\pm$0.04} \\
PCA (4D) &(rbf) & ImageNet &0.97$\pm$0.05 &0.93$\pm$0.05	&0.95$\pm$0.04 & \\
\cline{3-7}
& & COCO &0.91$\pm$0.07 &0.90$\pm$0.08	&0.90$\pm$0.07	&\textbf{0.90$\pm$0.07} \\
& & SUN &0.90$\pm$0.07 &0.91$\pm$0.08	&0.90$\pm$0.07 & \\
\cline{3-7}
& & ImageNet &0.93$\pm$0.09 &0.90$\pm$0.08	&0.91$\pm$0.07	&\textbf{0.91$\pm$0.07} \\
& & SUN &0.90$\pm$0.08 &0.92$\pm$0.09	&0.91$\pm$0.07 & \\
\hline
\end{tabular}
\caption{\label{tab:table1} Results demonstrating mean classification performance measures using ML classifiers that yield the best performance for different types of feature sets across all subjects and all sessions. As seen from the table, the best classification result is obtained with autoencoder-driven 4D PCA features for classifying visual networks of COCO, ImageNet, and SUN into corresponding datasets.}
\end{table}

\section*{Discussion}
Machine vision has undergone a revolutionary transformation with the introduction of large-scale image datasets and statistical learning approaches. However, the study of human visual perception through neuroimaging remains an open fundamental challenge. 
The present study aims to understand if the temporal profile of brain activity is distinct when individuals view images like a ``single focused object" versus ``a complex outdoor landscape". For example, networks processing complex scenes from the COCO dataset might exhibit different connectivity patterns compared to those processing object-centric images from ImageNet.
To address this, the study utilizes the ``BOLD5000" dataset which contains fMRI volumes acquired while viewing images from three diverse computer vision datasets, namely COCO, ImageNet, and SUN; serves as an excellent resource to investigate variations in brain activities related to vision. The above-listed computer vision datasets contain images that present visual information of varying complexities. For instance, ImageNet images typically depict singular-focused unoccluded objects, occupying a large portion of the image with uniform illumination. On the other hand, COCO images feature single or multiple objects commonly encountered in everyday life, appearing at various scales. SUN contains images of natural environments with cluttered backgrounds, varying illumination, and occlusions, where the focus is not on specific objects. The differences among these datasets have been previously explored in literature \cite{ref52,ref53}. 
The current study quantifies the differences in topological characteristics of visual networks of COCO, ImageNet, and SUN, as obtained from the correlation between the fMRI BOLD time series. Examining these differences in visual networks will capture how visual perception is influenced by the characteristics of the visual stimuli, conveying distinct information. Towards this, the classification of these visual networks as being associated with the corresponding visual datasets is performed based on their distinct topological characteristics. \par
The first contribution of this study is to incorporate partial correlation estimates along with marginal correlation to obtain a more robust and reliable visual network, built using true connectivity. The visual network is formed by considering only those network edges where both marginal and partial correlation estimates are positive. When both types of correlations are positive, it indicates that the two brain regions are simultaneously activated and have a direct positive association with each other, independent of the influence of other brain regions. Thus, by considering both marginal and partial correlation, spurious or indirect connections can be filtered out that might result from confounding factors, noise, or global signals. This enhances the reliability and accuracy of the visual network. Moreover, given that the BOLD signal is not a direct measurement of the electrical responses of the neurons, incorporating both marginal and partial correlation estimates ensures that the identified connections are more likely to reflect actual neural interactions related to visual stimuli.  \par
The second contribution of this study arises from effectively analyzing the differences in topological characteristics among the derived visual networks of COCO, ImageNet, and SUN. To achieve this, the topological persistence diagram of the visual network is computed for both dimension 0 and dimension 1.
Subsequently, K-means clustering (K=3, one cluster for highly
persistent descriptors, another for relatively less persistent descriptors, and yet another for spurious descriptors that include those with the lowest lifespan) is applied to each persistence diagram, enabling the extraction of various cluster features that reveal distinct topological patterns, corresponding to unique visual network structures associated with COCO, ImageNet, and SUN datasets. \par
As the third contribution of this study, we propose a novel deep hybrid learning model to classify visual networks of COCO, ImageNet and SUN using the extracted K-means-derived topological features. The deep hybrid learning model includes a single-layer autoencoder that takes the K-means-derived feature as input to produce a 4D latent space. PCA is applied to these 4D latent space features for traditional machine learning (ML) based classification. The results demonstrate that the proposed deep hybrid learning model achieves improved classification performance when compared to traditional ML classification using the following feature types: \textit{Type 1.} direct K-means features followed by PCA (without autoencoder), henceforth called ``K-means-derived PCA features", \textit{Type 2.} direct utilization of K-means features (without autoencoder) and \textit{Type 3.} autoencoder 4D latent space features (without PCA). 
However, performing classification directly on the K-means-derived PCA features (\textit{Type 1.}) is not optimal since PCA doesn't necessarily extract the most relevant topological features for the classification task. 
%Hence, the result of classification using K-means-derived PCA features is not tabulated in Table~\ref{tab:table1}. 
%It focuses on finding orthogonal directions of maximum variance in the data. These directions may not always correspond to the most discriminative features for classification. 
To further explain, let us examine two cases that arise from applying PCA: (A) selective low-dimensional feature space, (B) exhaustive high-dimensional feature space. In the case of selective low-dimensional feature space, the hypothesis is that the discriminative patterns present in the data may not be captured completely. This could lead to sub-optimal performance.
On the other hand, in the case of exhaustive high-dimensional feature space, it is seen that the number of instances may be insufficient (as in our study) to span the feature space. This would lead to poor generalization capability and hence reduced performance.
Moreover, as the underlying data distribution of this study is non-linear and the classes cannot be assumed to be linearly separable in the original feature space, utilizing PCA (a linear transformation technique) alone on the original feature space may not be effective in achieving good classification performance. To overcome this limitation, we propose a new deep hybrid architecture in the current study. \par
The proposed deep-hybrid architecture offers the following advantage: the unsupervised learning of the autoencoder's latent space is designed to capture the most essential information in the data for accurately reconstructing the original data. The autoencoder learns a compact representation of the data by compressing it into a lower-dimensional 4D latent space. This latent space contains features that are highly informative for the reconstruction task, effectively encoding the salient characteristics of the input data. PCA identifies the directions of maximum variance in the latent space representation. Thus, applying PCA to this already informative 4D latent space further refines the latent space embeddings by providing a new uncorrelated compact representation.    
Overall, the deep-hybrid architecture leverages the strengths of both the autoencoder and PCA to create a refined, informative, and low-dimensional feature space for better discriminability. This finally leads to improved classification performance by mitigating overfitting and allowing better generalization to new, unseen data. This makes it a compelling approach for ML-based classification tasks, especially when dealing with complex and high-dimensional datasets with limited observations.\par
While the study allows for a comprehensive exploration of how visual network topology differs in response to a wide range of visual stimuli with varying complexities, it is essential to acknowledge the limitations of this study. The first limitation of this study is the unavailability of sufficient subjects. BOLD5000 dataset includes only 4 subjects, which makes it difficult to draw conclusive inferences. To make this study more applicable to a wider context, we would need a larger group of participants, more fMRI sessions, and more diversity of stimulated images. Even though 5,254 images with various complexities provide a substantial dataset for studying brain visual responses with fMRI, it's still smaller when compared to human visual experience in everyday life. Thus, further studies are needed to perform for generalize the results to other visual datasets or real-world visual scenarios. Secondly, although brain network connectivity analysis provides valuable insights into brain network organization, the results of the current study are based on partial correlation and do not necessarily imply causal relationships. Thus, future studies should incorporate techniques such as Granger causality or dynamic causal modeling to identify directional influences between brain regions during visual processing. Thirdly, to gain a deeper understanding of how brain connectivity patterns relate to visual stimuli, future studies must analyze behavioral data which is lacking in the BOLD5000 dataset. This integration can help correlate brain activity with behavioral performance and provide insights into the cognitive processes underlying visual perception.  Addressing the limitations and exploring these potential avenues can lead to more robust and meaningful insights into the brain's functional organization in relation to diverse visual information. Nevertheless, the results obtained from this study provide valuable insights and showed a strong starting point for future exploration into how the brain processes visual information. 

\section*{Conclusion}
The work described in this paper presents a novel approach to classify visual networks across visual datasets (COCO, ImageNet, and SUN) having diverse image complexities using fMRI BOLD time series. Being able to classify these stimuli based on the BOLD time series provides insights into how different visual stimuli from COCO, ImageNet, and SUN datasets may lead to distinct brain network connectivity patterns. 
In this study,the differences in connectivity patterns of visual networks of COCO, ImageNet, and SUN are quantified by examining differences in network topological characteristics using persistence diagrams and its K-means clustered features. A novel deep-hybrid model is introduced for classifying visual networks into corresponding visual datasets of COCO, ImageNet and SUN. By identifying the distinctions using a combination of TDA of visual networks with deep-hybrid learning strategy, this study significantly contributes to the advancement of our knowledge on visual perception of static images. Comparing the visual networks of COCO, ImageNet and SUN reveal how human brain processes different types of visual information with varying object scales, contexts and complexities. This knowledge can be valuable for understanding the cognitive processes involved in object recognition, scene perception and visual memory. Moreover, the distinctive topological characteristics of visual networks associated with each dataset, as revealed from this study, could potentially lead to the development of future neuroimaging biomarkers for diagnosing visual processing disorders such as visual agnosia or prosopagnosia and also to track changes in visual cognition over time.

%%%%%%%%%%%%%%%%%%%%%%%%%%%%%%%%%%%%%%%%%%%%%%%%%%%%%%%%%
\section*{Proposed Methodology}
In this paper, the fMRI time series, which is the representation of BOLD intensity distribution over time, is utilized to study the differences in topological characteristics of three visual networks, constructed while viewing images from COCO, ImageNet, and SUN. 
Different software and libraries are used to conduct the entire research work. The code for constructing visual networks from the fMRI time series is executed in MATLAB (MathWorks - version R2018a) and runs on a machine with an Intel-R Core I3 4005U CPU 1.70 GHz processor with 4.0 GB RAM. For computing the topological descriptors using the persistence diagram we use the GUDHI \cite{gudhi:urm} library. For classification using the proposed deep hybrid learning model, colab notebooks and keras are used. Colab provides 16GB of NVIDIA Tesla T4 GPU.

\subsection*{Time series extraction from fMRI image volume}
\label{sec:ts_extraction}
\begin{figure}[ht]
\centering
\includegraphics[width=0.98\linewidth]{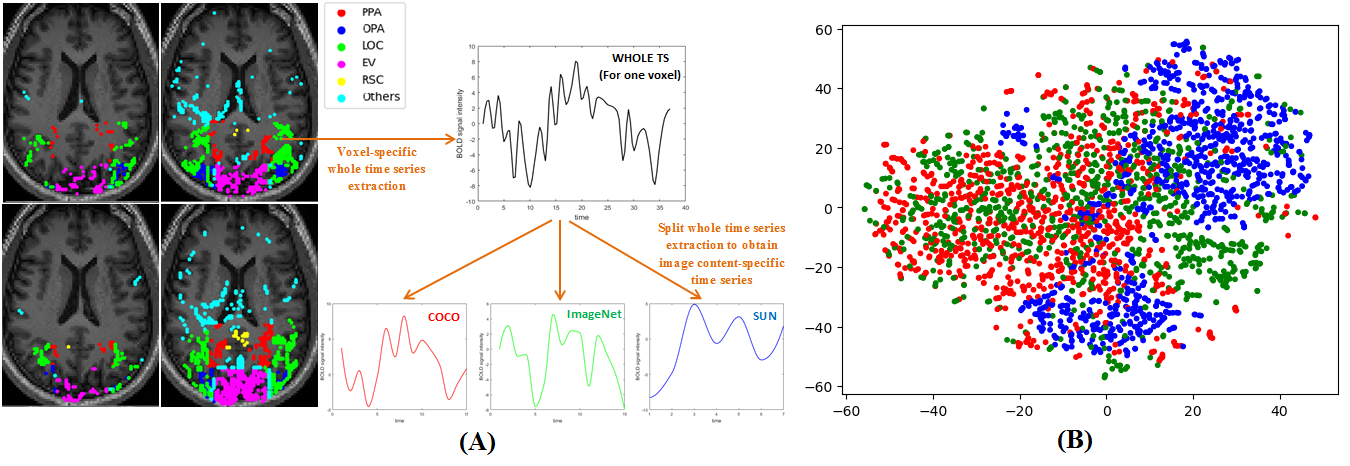}
\caption{The MRI images as shown in Figure (A) (\textit{left}) illustrate voxel activation within the brain while viewing images, for all four subjects. Figure (A) (\textit{Right}) shows one example of whole time series \textit{Black} which is extracted from one active voxel for one representative subject. Image complexity-specific time series is obtained by splitting this whole time series according to the BOLD intensity values of distinct image categories (COCO, ImageNet, and SUN) as seen by the participants at each time stamp. These three sets of time series- representing COCO, ImageNet, and SUN, as obtained from the whole time series are shown in \textit{Red}, \textit{Green}, and \textit{Blue} color respectively. Figure (B) shows the t-distributed stochastic neighbor embedding (t-SNE) visualization of these time series as obtained from each and every active voxel of COCO, ImageNet, and SUN for one representative subject.}
\label{fig:ts-tsne}
\end{figure}

In order to extract the whole fMRI time series, first, it is necessary to have information about active voxel locations in the brain. In our study, SPM \cite{ref15} toolbox was used to get voxel locations from 4D ($x,y,z$ and $t$) fMRI data which were found to be significantly active during each run of the experiment. The locations of active voxels are mapped on MRI images of corresponding subjects and shown in Figure~\ref{fig:ts-tsne}(A). As seen from the figure, these locations of active voxels are found within $5$ visual ROIs as defined in \cite{ref14} which are the parahippocampal place area, the retrosplenial complex, the occipital place area, the early visual area, and lateral occipital complex. Some voxels that are found to be activated in other sub-cortical regions are labeled as ``others" in our study. From each run, the whole time series of length $37$ (since, the no. of stimuli $= 37$, at each run) is extracted for each of the active voxels. The obtained voxel-specific time series is detrended and the Z-score is normalized before it is used for further analysis. \par
In order to obtain the image complexity-specific fMRI time series, the whole time series is split into three parts according to the BOLD intensity values of images that represent a particular visual dataset class; i.e. whether it belongs to COCO or ImageNet or SUN (Figure~\ref{fig:ts-tsne}). Thus, each of these three time series illustrates the BOLD intensity distribution while viewing images from COCO, ImageNet, and SUN. It is to be noted that, the length of these three time series corresponding to these three visual datasets are not the same due to randomness in the total number of image presentations to the participants from these three visual datasets.

%%%%%%%%%%%%%%%%%%%%%%%%%%%%%%%%%%%%%%%%%%%%%%%%%%%%%%%%%

\subsection*{Estimating marginal and partial correlation between BOLD time series}
Correlation analysis is the classical technique that has been widely used to study the nature of connectivity between two brain regions\cite{ref1,ref16,ref17,ref22}. It is seen that most of the studies used Pearson correlation (also known as marginal correlation), which only captures the marginal association between network nodes. However, brain connectivity analysis using only Pearson correlation is not sufficient since it fails to capture the direct or true connectivity if exists between network nodes. For example, significant correlations between two nodes, say, X and Y may occur due to their common connections to a third node, say Z, even if X and Y are not directly connected \cite{ref16,ref17}. Thus using marginal correlation, it is hard to differentiate the network edges that reflect true connectivity from the edges that reflect the connectivity caused by confounders. The statistical technique that has shown great potential in addressing this major issue is partial correlation \cite{ref16,ref17,ref18}. \par
Partial correlation estimates correlations after regressing out spurious effects from all the other nodes in the network; making it a true measure of network connectivity. Thus, a zero value in partial correlation suggests an absence of direct connectivity between network node pairs. Strong evidence from the literature suggests that partial correlation is one of the top techniques that outperform the traditional techniques and showed high sensitivity to find true functional connectivity between network nodes \cite{ref16,ref17,ref19,ref43,ref44,ref45,ref46}. \par

\subsubsection*{Partial correlation: Proof-of-the-concept}
\begin{figure}[ht]
\centering
\includegraphics[width=0.95\linewidth]{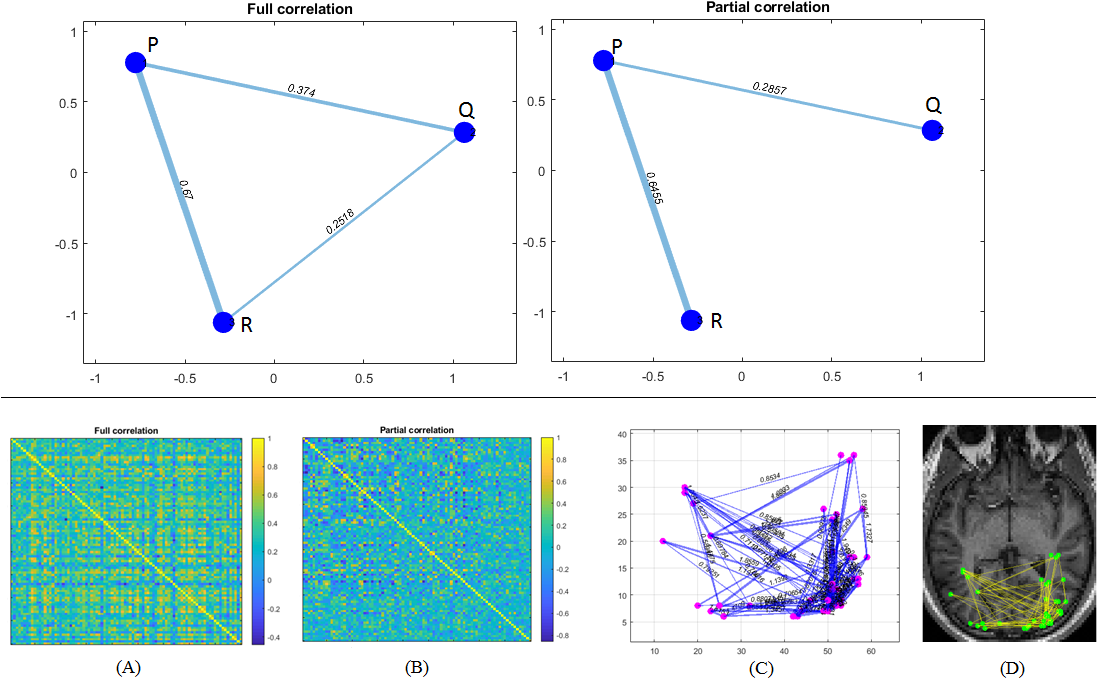}
\caption{\textit{Top-row:} Toy example demonstrating marginal correlation (Pearson's r) (\textit{left}) and partial correlation (\textit{right}) for a 3-node network. Visual network connectivity is computed using both of these correlation measures. As seen from the figure, partial correlation can detect the conditional independence between node $Q$ and node $R$, whereas Pearson's correlation only measures the marginal association between node $Q$ and node $R$ which is resulted due to their common connection with node $P$. 
\textit{Bottom-row:} Marginal and partial correlation estimates as obtained for one fMRI session, for one representative subject is shown in Figure (A) and Figure (B), respectively. A lot of positive connectivity was seen to be either removed or lessened in strength in partial correlation, as expected since partial correlation captures the direct linear association between network nodes after eliminating the spurious effects from all confounders. Figure (C) shows the graphical representation of visual networks with positive edge strengths as obtained from correlation analysis. The respective mapping of the network on the fMRI image is shown in Figure (D).}
\label{fig:toycorr}
\end{figure}

A toy example is shown in Figure~\ref{fig:toycorr} (\textit{Top-row}), demonstrating the difference between partial correlation and marginal correlation of a network having 3 nodes $P$, $Q$, and $R$. Let,
\[P \sim \mathcal{N}(\mu,\,\sigma^{2})\,.\]
\[Q = a_{1}P+\epsilon_{1}\]
\[R = a_{2}P+\epsilon_{2}\]
The time series generated from $P$, $Q$, and $R$ were used to compute partial correlation and marginal correlation, with $a_{1}=0.4$ and $a_{2}=0.9$. The result is presented in Figure~\ref{fig:toycorr}. Although both correlations were able to find the connectivity between $P$ and $Q$, and between $P$ and $R$, unlike marginal correlation, no direct association was captured between $Q$ and $R$ in partial correlation (correlation value $\approx$ $0$). Hence it is concluded that the association between $Q$ and $R$ that was observed in marginal correlation, caused due to their common connection with node $P$. This proves the ability of partial correlation to eliminate spurious correlations, caused by the presence of confounders, and hence provides a more reliable estimate for measuring direct connectivity in brain networks.
Due to the short duration of the BOLD hemodynamic response, in this study, the marginal correlation of two time series is computed only at zero-lag \cite{ref2,ref20,ref21}.

Consider, we have $N$ distinct brain locations that define brain activity which is measured as fMRI BOLD time series at $M$ time points. In a Gaussian model, let us define the BOLD time series from $N$ regions ($n = 1,2,\ldots, N$) across $M$ time points as $X_{n}=[x_{n1}\, x_{n2}\, \ldots \,x_{nM}]$ for $n = 1 \text{ to }N$, are independent and identically distributed as a multivariate
Gaussian $\mathcal{N}(\mu,\Sigma)$ where $\mu\in\mathbb{R}^{N}$ and $\Sigma_{N \times N}$ is the covariance matrix. Then, the marginal correlation of any two time series signals for the corresponding two distinct brain locations, ($i,j$) ($i, j$ varies from $1 $to $N$; $i \neq j$) can be computed as,
\begin{equation}
r_{ij}=\frac{cov(X_{i},X_{j})}{\sigma(X_{i})\sigma(X_{j})}.
\end{equation}

Partial correlation of two time series between region `$i$' and region `$j$' at t is defined as the correlation between $X_{i}$ and $X_{j}$ conditioning on all the other nodes \cite{ref16}, i.e.:\\
\[ \rho_{ij} = corr(X_{i}, X_{j}\mid X_{ij}), \text{ where } X_{ij}=\{X_{k}: 1\leq k\leq N;\; k \neq i \neq j \}\]

The interpretation of the values of $\rho_{ij}$ not equal to zero is the same as that with marginal correlation. However When $\rho_{ij}=0$, node-$i$ and $j$ becomes conditionally independent given the other nodes. 
Let us consider the case of a simple three-node network $P$, $Q$, and $R$ as shown in Figure~\ref{fig:toycorr}. The partial correlation between $Q$ and $R$ is derived as follows. First regressing the time series of $P$ against the time series of $Q$ and denoting the residual as $\xi_{1\mid3}$. Then, regress the time series of $Q$ against the time series of $R$ and denote the residual as $\xi_{2\mid3}$. The partial correlation between $Q$ and $R$ can then be obtained as the correlation between $\xi_{1\mid3}$ and $\xi_{2\mid3}$. Thus, when the number of nodes in the network exceeds the value $3$ (i.e. for $N\geq3$), the partial correlation of two nodes is defined as "the correlation of residuals obtained from linear regression of first node with all other confounders and linear regression of second node with all other confounders" \cite{ref16}.\par

However, the computation of partial correlation using the regression approach is not appropriate in terms of computational time. Moreover, it often fails if multicollinearity exists among time series. In such cases, instead of using linear regression, partial correlation is estimated effectively from the inverse covariance matrix, also referred as the precision matrix. 
Let, $\Sigma$ be the $N\times N$ covariance matrix of two time series, and let, $\Omega_{N\times N}$ = $\Sigma^{-1}$ = $(\omega_{ij})_{N\times N}$ be the precision matrix, then, the partial correlation between the time series of node-$i$ and node-$j$ can be derived from the precision matrix as
\begin{equation}
    \rho_{ij}=\frac{-\omega_{ij}}{\sqrt{\omega_{ii}\omega_{jj}}}
\end{equation}

Clearly, estimating partial correlation using a precision matrix requires a covariance matrix to be invertible. In neuroimaging applications, this is challenging when the number of active voxels exceeds the number of observations (less number of fMRI scans). In such cases, derivation of the precision matrix needs more computational load and it may not be stable. There are many algorithms proposed in the literature to overcome this difficulty \cite{ref16}. One such solution is to apply the Moore-Penrose pseudo-inverse of the covariance matrix to directly estimate the precision matrix. In our study, the Moore-Penrose pseudo inverse is used to estimate partial correlation.

\subsection*{Constrution of visual network}
\label{sec:Constrution of visual network}
A network is a collection of nodes (vertices) and links (edges) between pairs of nodes. In the context of brain networks, nodes usually represent distinct brain regions, while the edges represent connectivity strength between nodes\cite{ref2,ref22}. Thus, network connectivity can be mathematically represented as an undirected graph $G=(V,\, E)$, where, $V$ is a set of nodes and $E$ is a set of edges or links. Each link $l_{ij}\in E$, connecting a node-$i$ with a node-$j$, is associated with a weight (either positive or negative) which represents BOLD temporal correlations in brain activity that occur between two regions. Each network can be represented by a network connectivity matrix (also known as adjacency matrix) which is a symmetric matrix of size $N \times N$, where, $N$ is the number of nodes. Each $(i,j)$-th entry in this matrix denotes the strength of the edge connected between nodes `$i$' and `$j$'. 
In our study, the edge strength entries are estimated using both marginal and partial correlation. \par
Since correlation values vary from negative ($-1$) to positive ($+1$), the network constructed using correlation as connectivity measures can be categorized into two types: (i) positively-correlated network having positive edge strength and (ii) negatively-correlated network having negative edge strength. In the context of brain connectivity analysis, a positive correlation indicates that when brain activity in one region increases, increased activity is seen in the other region too; whereas a negative correlation indicates that when one brain region is more active, the activity in the other region decreases.
However, in the context of network physiology, analyzing brain networks with negative connectivity is still less understood and has been a subject of debate since several studies demonstrated that the negative correlation could be a result of artifact introduced by a global signal regression procedure during pre-processing of fMRI data \cite{ref31,ref32,ref33}; thereby needs to be meticulously interpreted.
Therefore, the current study utilizes only positively correlated networks for subsequent analysis. Visual networks are constructed when both marginal and partial correlation edge strength values exhibit a positive association.
The visualization of a positively correlated network for one subject is shown in the bottom row of Figure~\ref{fig:toycorr}.

%\begin{figure}[ht]
%\centering
%\includegraphics[width=0.95\linewidth]{images/corr-vbn.png}
%\caption{Illustration of Marginal correlation and Partial correlation as obtained for one representative subject, for one session is shown in Figure (A) and Figure (B), respectively. A lot of positive connectivity was seen to be either removed or lessened in strength in partial correlation, as expected, since partial correlation captures the direct linear association between network nodes after eliminating the spurious effects from all confounders. Figure (C) shows the graphical representation of visual networks with positive edge strengths as obtained from correlation analysis. The respective mapping of network on fMRI image is shown in Figure (D).}
%\label{fig:visual network}
%\end{figure}

\subsection*{Topological data analysis of visual networks} 
%The differences in the topological characteristics of the obtained visual networks are analysed by extracting relevant topological descriptors from the persistence diagram. 
In the current research, we employ tools from TDA to capture topological features from the constructed visual networks and to examine the distinctive characteristics present in the visual networks corresponding to  COCO, ImageNet, and SUN, obtained for each fMRI session. In computational topology, Betti numbers are important topological features to distinguish two graphs or networks.
The $0$-th Betti number of a graph represents the count of connected components and the $1$-st Betti number counts the number of independent loops which are the dimensions of the $0$-th and $1$-st homology groups, respectively. However, in the current study, to analyze more significant topological features, we consider the persistent homological features corresponding to a graph filtration of each visual network by computing its persistence diagram. 
The subsequent sub-sections provide a brief description of the tools from the TDA used in the current study.

\subsubsection*{Graph Filtration}
%For each $r \in \R$, we define the subgraph $G_{\leq r}$ of $G$ as follows.
%\begin{itemize}
 %   \item An edge $e \in G_{r}$ if and only if $W(e) \leq r$.
  %  \item A vertex $v \in G_{r}$ if and at least one its adjacent edge $e$ satisfies $W(e) \leq r$.
%\end{itemize}
%The criteria for vertices and edges in $G_{\geq r}$ is similar to those for $G_{\leq r}$ with $\geq$ in the place of $\leq$.
Consider a weighted graph $G=(V,\, E)$, where $V$ and $E$ represent the sets of vertices and edges, respectively. The weight of an edge $e$ is denoted by $w(e)$. A filter function $f: G \rightarrow \R$, on $G$, is defined as follows:
\begin{itemize}
    \item for an edge $e \in E$, the value of $f(e)$ is set to $w(e)$,
    \item for a vertex $v \in V$, the value of $f(v)$ is obtained by selecting the minimum among all edge weights associated with the edges incident on $v$.
\end{itemize} 

\noindent%Once the filtration for a graph is established, we can compute persistent topological features in them. 
For each $r \in \R$,  the subgraph $G_{\leq r}$ of $G$ is defined as the collection of vertices and edges in $G$ with $f$-values at most $r$. Similarly, the subgraph $G_{\geq r}$ consists of vertices and edges with $f$-values greater than or equal to $r$. Let $w_1\leq w_2\leq  \ldots\leq w_{|E|}$ be the weights of the edges in $G$, in non-decreasing order, where $|E|$ denotes the number of edges in $G$. This arrangement yields two sequences of graphs, each referred to as a \emph{filtration}. The sequence $\emptyset = G_{\leq 0} \subseteq G_{\leq w_1} \subseteq G_{\leq w_2} \subseteq \ldots \subseteq G_{\leq w_n} = G$ is called the sublevel set filtration of $f$ and  the sequence $G_{\geq w_n} \subseteq G_{\geq w_{n-1}} \subseteq \ldots \subseteq G_{\geq w_{0}} = G$ is referred to as the superlevel set filtration of $f$. \figref{fig:GraphFiltration-PersistenceDiagrams}(b) illustrates the sublevel set filtration associated with a graph, highlighting the sequence of subgraphs as the filtration progresses. The filtrations provide a basis for computing the persistent topological features that exist within the graph. The persistence of topological features is encoded in a descriptor known as the \emph{persistence diagram}. In the next section, we provide a brief description of the persistence diagram, discuss its applications and an intuition of how to construct a persistence diagram based on graph filtration.

\subsubsection*{Persistence Diagrams}
Persistent homology \cite{2002-Edelsbrunner-Persistence, 2005-Zomorodian-ComputingPersistentHomology} is a fundamental tool in TDA and stands as one of the early applications of algebraic topology in data analysis \cite{2009-Carlsson-Topology-and-Data}. By leveraging on a robust mathematical theory and exhibiting stability in the presence of minor noise disruptions\cite{2007-Cohen-Steiner-Bottleneck}, persistent homology has introduced a novel qualitative and quantitative descriptor known as the persistence diagram. This descriptor encodes the persistence features in data as a collection of points in the two-dimensional Euclidean space $\mathbb{R}^2$. 
\begin{figure}[h]
\centering
\includegraphics[width=0.98\linewidth]{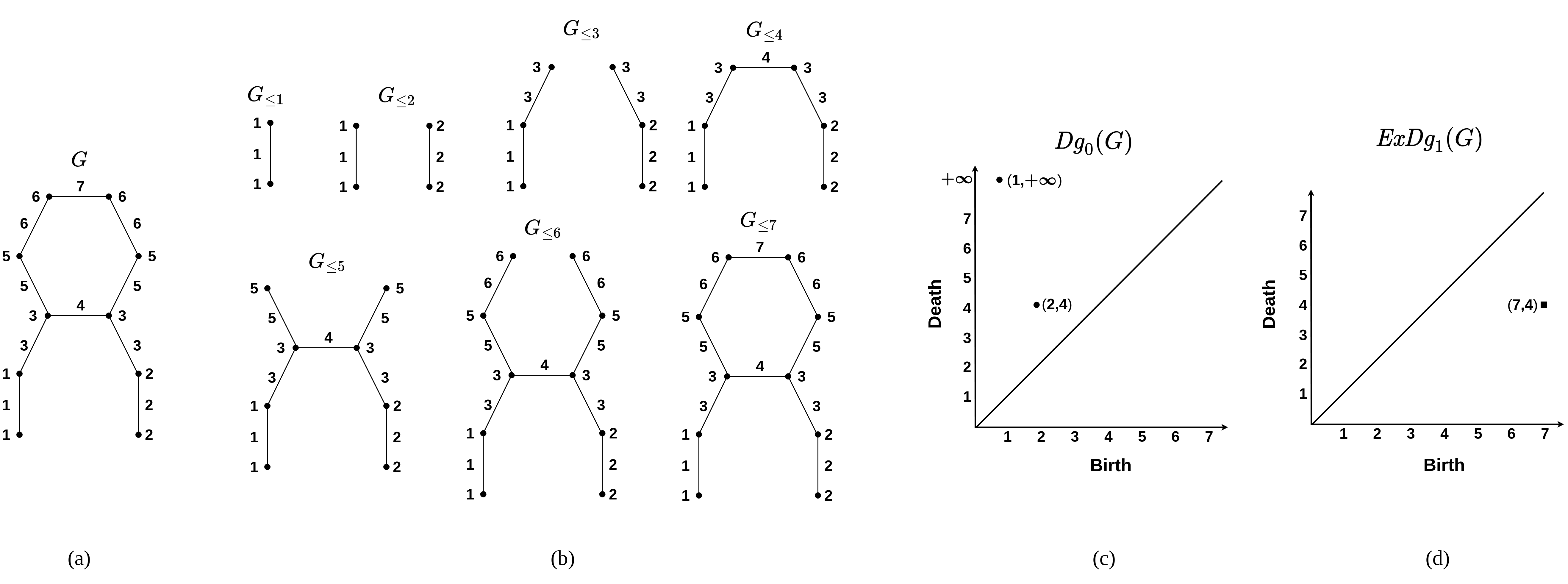}
\caption{Construction of persistence diagrams from a graph. (a) A graph $G$. (b) Sublevel set filtration corresponding to $G$ based on a filter function $f:G \rightarrow \R$. The values of $f$ are indicated at the vertices and edges. (c) $0$-th ordinary persistence diagram ($\Dg_0(G)$). (d) $1$-st extended persistence diagram ($\ExDg_1(G)$). The points in ordinary and extended persistence diagrams are denoted by circular points and squares, respectively. At $G_{\leq 4}$ in the sublevel set filtration, a component born at $G_{\leq 2}$ merges with another component born at $G_{\leq 1}$. This feature is captured by the point $(2,4)$ in $\Dg_0(G)$. Moreover, the component born at $G_{\leq 1}$ persists throughout the filtration and is indicated by the point $(1,\infty)$ in $\Dg_0(G)$. Finally, the loop in $G$ is encoded by the point $(7, 4)$ in $\ExDg_1(G)$.}
\label{fig:GraphFiltration-PersistenceDiagrams}
\end{figure}

\begin{figure}[ht]
\centering
\includegraphics[width=0.98\linewidth]{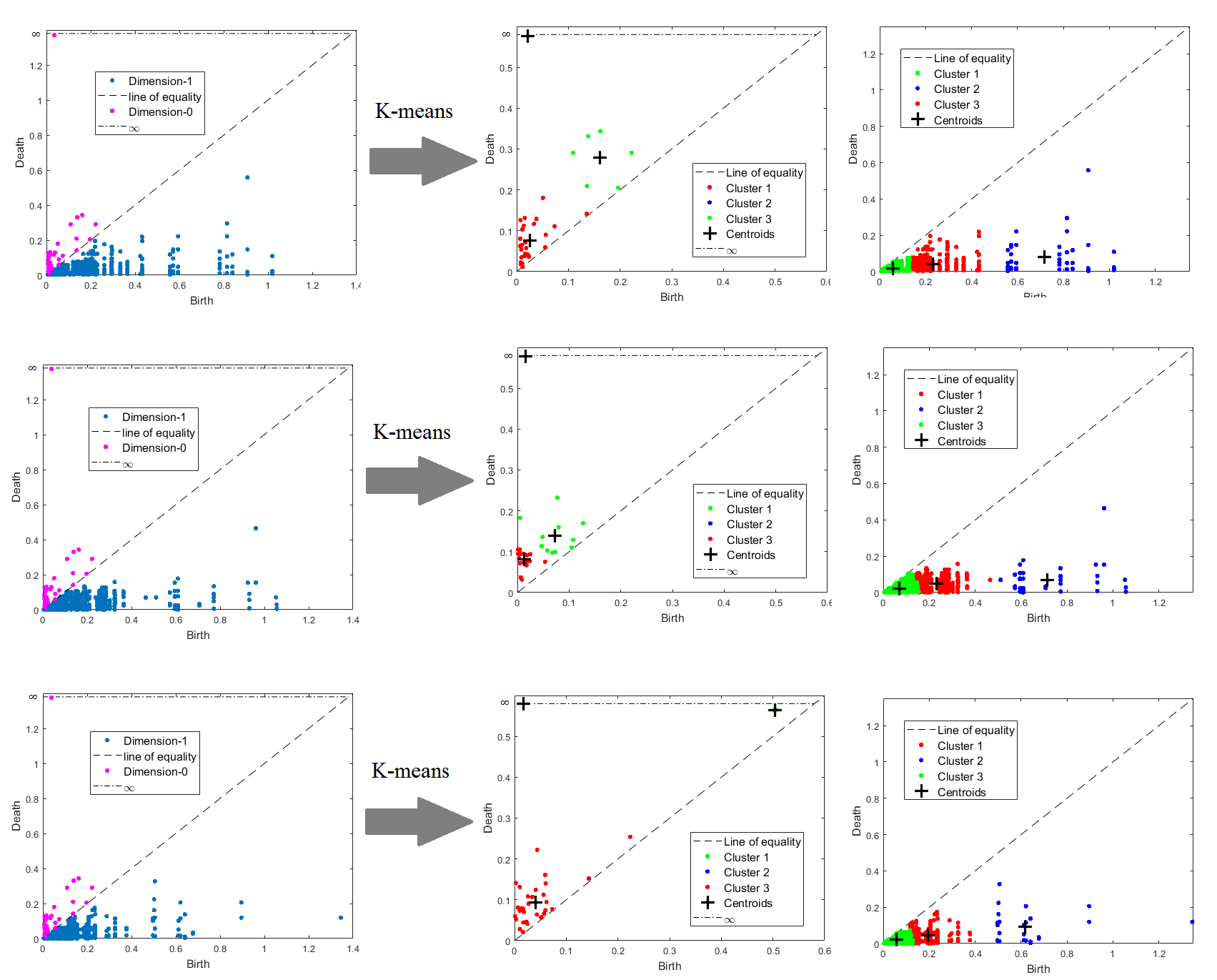}
\caption{Illustration of K-means clustering with \textit{K=3}. \textit{First column:} shows persistence diagram plot of visual network for COCO (\textit{first row}), ImageNet (\textit{second row}) and SUN (\textit{third row}), as obtained for one particular fMRI session of a representative subject. The 0- and 1-dimensional persistence diagrams are shown in color \textit{Pink} and \textit{Blue}, respectively. The \textit{Second column:} and \textit{Third column:} illustrate the results of K-means clustering, executed on 0-dimensional and 1-dimensional persistence diagram, respectively.}
\label{fig:kmresult}
\end{figure}

Persistence diagrams have found applications in a wide range of fields, including endoscopy analysis \cite{2016-Dunaeva-Endoscopy-Analysis}, pulmonary diseases \cite{2018-Brodzki-Pulmonary-Disease}, finance \cite{2017-Gidea-TDA}, shape-matching \cite{2022-Ramamurthi-Shape-Matching} and fingerprint classification \cite{2019-Giansiracusa-Persistence-FingerprintClassification}, among others. Notably, persistent homology has been utilized in the analysis of brain networks for differentiating various types of brain disorders \cite{2011-Lee-Persistent-Homology-Brain-Networks, 2012-Lee-Persistent-Homology-Brain-Networks}, detecting transition between states in EEG (Epileptiform) signals \cite{2016-Merelli-EEG}, distinguishing schizophrenia patients from healthy individuals \cite{2021-Stolz-schizophrenia-experiments}, identifying epileptic seizures \cite{2021-Caputi-Epilepsy} and distinguishing between male and female brain networks \cite{2022-Das-TDA-Brain-Networks}. In this study, we exploit the information encoded in persistence diagram to analyze the connectivity of visual networks and apply it for categorizing visual networks across visual datasets. Here, we provide an intuitive explanation of how persistence diagrams are constructed based on graph filtrations. Further details on persistence diagram can be found in literature \cite{2010-Edelsbrunner-book}.
%For further details, we refer readers to the book \cite{2010-Edelsbrunner-book}.

Let us consider a graph $G$. To capture the $0$-dimensional homological features, the birth and death of connected components are examined in the sublevel set filtration. In computational topology, connected components are called $0$-dimensional homological features. Hence, the persistence diagram that encodes the birth and death of connected components is called the $0$-th ordinary persistence diagram, denoted by $Dg_0(G)$. Mathematically, the points in $Dg_0(G)$ are computed from the birth and death of homology classes in a sequence of homology groups corresponding to the sublevel set filtration \cite{2010-Edelsbrunner-book}. Note that a component is considered to be born at $G_{\leq b}$ if it exists in $G_{\leq b}$ but not in $G_{\leq a}$ for any $a < b$. Let $C_1$ and $C_2$ be two components in the sublevel set filtration born at $G_{\leq c}$ and $G_{\leq d}$, respectively, with $c < d$. If $C_1$ and $C_2$ merge at $G_{\leq e}$, then $C_2$ is said to die at $G_{\leq e}$ (by the \emph{elder rule} of persistence homology \cite{2010-Edelsbrunner-book}). This is represented by the point $(d,e)$ in $\Dg_0(G)$. If a component is born at $G_{\leq a}$ and never dies throughout the filtration, it is represented by the point $(a,\infty)$ in $\Dg_0(G)$. Note that each point in $\Dg_0(G)$ occurs above the diagonal since it corresponds to a connected component in the sublevel set filtration which is born at $G_{\leq a}$ and dies at $G_{\leq b}$ with $a < b$. \figref{fig:GraphFiltration-PersistenceDiagrams}(c) displays the $0$-th ordinary persistence diagram corresponding to a sublevel set filtration. 
%A component born at $G_{\leq 2}$ merges with another component born at $G_{\leq 1}$. This merging event happens at $G_{\leq 4}$ and is represented by the point $(2,4)$. Additionally, the component born at $G_{\leq 1}$ never dies, which is indicated by the point $(1,\infty)$. % Since $Dg_0(G)$ encodes the birth and death of $0$-dimensional homological features, it is referred to as the $0$-dimensional persistence diagram. We now describe the birth and death of connected components.

Next, we discuss capturing the loops in the graph which are the $1$-dimensional homological features in a persistence diagram. To encode loops, one considers the birth and death of homology classes in a filtration obtained by combining the sublevel set and superlevel set filtrations, called extended filtration. Therefore, the persistence diagram encoding the loops of $G$ is called the $1$-st extended persistence diagram and is denoted by $\ExDg_1(G)$. To construct $\ExDg_1(G)$, one considers the $1$-dimensional homology classes that persist throughout the sequence of absolute homology groups corresponding to the sublevel set filtration. Such homology classes are called essential homology classes. Then the deaths of these essential homology classes in the relative homology groups in the superlevel set filtration are considered. Note that an essential homology class which is born at the sublevel set filtration and dies at the superlevel set filtration is encoded as a point in $\ExDg_1(G)$. Specifically, for a loop $l$ in $G$, let $a$ and $b$ be the minimum and maximum values of $f$ along the edges of $l$. Then, the loop is captured by an essential homology class which is born at $G_{\leq b}$ in the sublevel set filtration and dies at $G_{\geq a}$ in the superlevel set filtration. This feature is encoded as the point $(b, a)$ in $\ExDg_1(G)$. We note, contrary to $\Dg_0(G)$, in $\ExDg_1(G)$, the birth of a loop is essentially greater than its death in the extended filtration and hence the points in $\ExDg_1(G)$ occur below the diagonal. \figref{fig:GraphFiltration-PersistenceDiagrams}(d) shows the $1$-st extended persistence diagram of a graph. In this paper, we compute $\Dg_0$ and $\ExDg_1$ corresponding to the visual networks, representing visual datasets - COCO, ImageNet, and SUN and utilize them for the purpose of classifying these visual networks across visual datasets. 
% Mathematically, a sequence of absloute homology groups going up in the sublevels set filtraion followed by a sequence of relative homology groups coming down in the superlevelset filtraion is considered, to capture the birth-death of these features. We provide a brief description of encoding loops and refer the reader to \cite{2010-Edelsbrunner-book} for more details. In computational topology, this filtration corresponding to a sequence of homological groups corresponding to the sublevelset filration followed by a sequence of relative homological groups corresponding to a superlevelset filtraion,

%To capture the $1$-dimensional features (loops), we compute the $1$-st extended persistence diagram, denoted as $\ExDg_1(G)$. This diagram incorporates information from both the sublevel set and superlevel set filtrations \cite{2010-Edelsbrunner-book}. For a loop $l$ in $G$, let $a$ and $b$ be the minimum and maximum values of $f$ along the edges of $l$. The loop is then captured by the point $(b, a)$ in $\ExDg_1(G)$. \figref{fig:GraphFiltration-PersistenceDiagrams}(d) shows the $1$-st extended persistence diagram of a graph. In this paper, we compute $\Dg_0$ and $\ExDg_1$ corresponding to brain networks and utilize them in the categorization of images.

\begin{figure}[ht]
\centering
\includegraphics[width=0.98\linewidth]{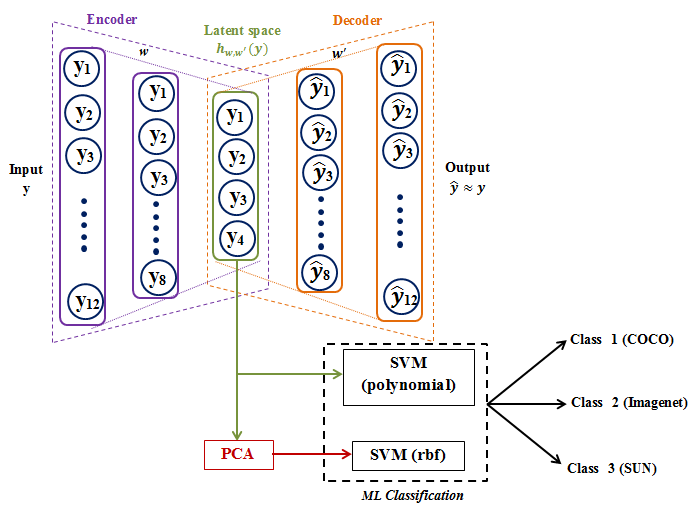}
\caption{The proposed deep-hybrid model for classifying visual networks across visual datasets of COCO, Imagenet and SUN is shown. The extracted K-means cluster features as obtained from the persistence diagram of each visual networks is given as input to the autoencoder. SVM classification is performed on both autoencoder latent space features as well as on PCA-reduced autoencoder latent space features.}
\label{fig:deep-hybrid model}
\end{figure}

\subsubsection*{Computation of topological descriptors from visual network}
The present study utilizes persistence diagram to derive relevant descriptors for quantifying the dissimilarities in the topological properties of visual networks, representing COCO, ImageNet, and SUN, as obtained from each fMRI session. These descriptors serve as powerful tools to capture and compare the distinguishing topological characteristics across these networks, offering valuable insights into their differences holistically.
As described in the previous subsection, persistence diagram captures the emergence and disappearance of topological features, such as connected components and loops. These significant events, known as ``birth" and ``death", are succinctly represented as points in the persistence diagram. The lifespan of each point $(b, d)$, which is born at the time `$b$' and dies at the time `$d$', is assigned a persistence value  $(d - b)$. Thus, in the persistence diagram, every point corresponds to a specific topological feature. The magnitude of the difference between the ``birth" and ``death" values indicates the feature's life-span or persistence. A larger difference implies a longer lifespan for the topological feature, making it more probable to be a significant characteristic of the underlying distribution from which the points are sampled. The resulting persistence diagram condenses these $(b, d)$ pairs into a 2D plot, where the x-axis represents the ``birth" and the y-axis represents the ``death" of the features. Figure~\ref{fig:kmresult} (\emph{first column}) illustrates persistence diagrams that summarize the topological descriptors of different visual networks. 
In this study, first, we compute the persistence diagram of each graph (visual network) corresponding to (i) COCO, (ii) ImageNet, and (iii) SUN, as constructed from each fMRI session.
Following this, we apply K-means clustering on each persistence diagram, to check the differences in cluster properties of point clouds in each persistence diagram, if any, among COCO, ImageNet, and SUN. This is described in the subsequent subsection and illustrated in Figure~\ref{fig:kmresult} \emph{second and third columns}. \par

\subsubsection*{Cluster analysis of persistence diagram and feature extraction}
Clustering (unsupervised learning) is a crucial tool utilized across various domains to gain insights and interpret data effectively. Among the popular approaches, K-means stands out as one of the most widely used methods. With a fixed number of clusters (K), the K-means algorithm aims to position cluster centers and define boundaries in a way that minimizes the sum of squared distances of points to the center within each cluster.
The idea behind directly applying K-means clustering on each persistence diagram is driven by the hypothesis that \textit{distinctive features that differentiate three different visual networks as obtained for COCO, ImageNet, and SUN datasets for a specific fMRI session, can be captured through the variation in cluster properties of point clouds of the corresponding persistence diagram.}
In this study, the value of `K' is set to 3, in order to form three distinct clusters from each persistence diagram. Each cluster represents a specific category of features with different lifespans. The first cluster contains less persistent features, those with the shortest lifespan. The second cluster encompasses highly persistent features, exhibiting the longest lifespan. Finally, the third cluster captures features with moderate persistence.  These moderate persistent features would fall into a cluster that is distinct from the cluster of short-lived features and the cluster of long-lived features. 
%These moderate features can contribute valuable insights that persist for a significant period without being as enduring as the high-persistent features nor as transient as the low-persistent ones. 
%The identification and analysis of moderate persistent features help to understand the intermediate patterns present in the data, shedding light on important characteristics that exhibit a moderate lifespan.
By employing K-means clustering, we gain valuable insights into the underlying cluster features across lifespans of different topological descriptors. Furthermore, this approach enables us to highlight the differences between distinct temporal characteristics of features related to visual networks of COCO, ImageNet, and SUN. As a result, we attain significant insights into the temporal evolution of these networks, enhancing our understanding of their dynamic behavior over time. Thus, K-means clustering analysis of the persistence diagram provides a comprehensive view of the data, shedding light on both its connectivity aspects and temporal dynamics within the context of brain network exploration. 

In this work, corresponding to a visual network $G$, six different K-means cluster features are computed based on cluster size and cluster centroid for each of the $0$- and $1$- dimensional persistence diagrams $\Dg_0(G)$ and $\ExDg_1(G)$, respectively. These are as follows: 
\begin{enumerate}[label=(\roman*)]
    \item  Fraction of points in less-persistent feature cluster.
\item  Fraction of points in moderately persistent feature cluster.
\item  Fraction of points in high-persistent feature cluster.
\item  Distance (perpendicular distance) of the less-persistent cluster centroid from the diagonal of the persistence diagram.
\item  Distance of the moderately persistent cluster centroid from the diagonal of the persistence diagram.
\item  Distance of the high-persistent cluster centroid from the diagonal of the persistence diagram.
\end{enumerate}
Thus, corresponding to each visual network, a $12$-dimensional feature vector is obtained by combining the features from the $0$- and $1$-dimensional persistence diagrams which are used for the classification of visual networks across visual datasets: COCO, ImageNet, and SUN.

\begin{figure}[ht]
\centering
\includegraphics[width=0.98\linewidth]{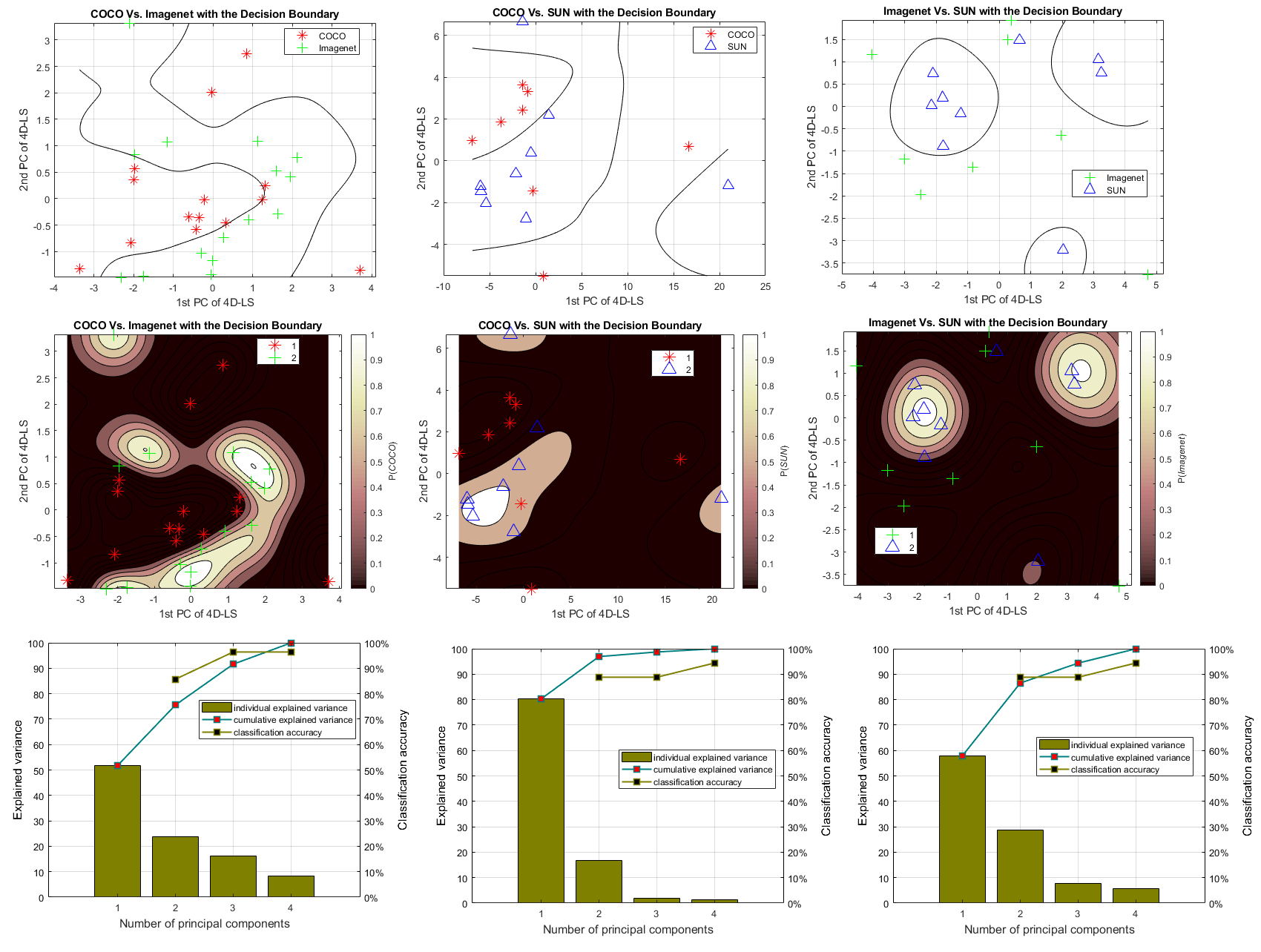}
\caption{Visualization of the performance of SVM-RBF model for classification of visual network of COCO vs. ImageNet (\textit{left}), COCO vs. SUN (\textit{middle}) and ImageNet vs. SUN (\textit{right}) for a representative subject. The \textit{top-row} shows the SVM decision boundary. 
The \textit{middle-row} shows the positive class posterior probability region to better understand how the SVM model separates different classes and identify areas of uncertainty where the model is less confident about its predictions.
The \textit{bottom-row} shows a Scree-Accuracy plot for the respective classification at different levels of explained variances, obtained from PCA. For all subjects, it is observed that the minimum top three principal components are required to achieve the minimum accuracy of 90\%, in classifying these visual networks across visual datasets.}
\label{fig:svm-all}
\end{figure}

\subsection*{Deep hybrid learning for classification}

This section describes a novel deep hybrid learning framework that we have designed for subject-specific classification among visual networks of COCO, ImageNet, and SUN across all sessions. The proposed framework combines deep learning with conventional ML algorithms. Several studies in the literature have already demonstrated the superior performance achieved by such deep hybrid approaches \cite{ref47,ref48,ref49,ref50,ref51}. In the proposed architecture, the extracted K-means features from the persistence diagrams are used as input to an autoencoder, allowing us to capture relevant information in its latent space. Following this, PCA is applied to the latent space embeddings for further refinement. Finally, these refined PCA-derived features are fed into the SVM classifier, resulting in higher accuracy in classifying the visual networks of COCO, ImageNet, and SUN across multiple fMRI sessions. The proposed deep hybrid learning model is illustrated in Figure~\ref{fig:deep-hybrid model} which has shown its potential in executing complex visual network classification tasks with high accuracy.

\subsubsection*{Input representation} 
For a specific subject, the features of size $m \times n$, as derived from K-means clustering from the persistence diagrams are fed as input to the autoencoder model. Here, the value of \textit{m} is $12$ which refers to the dimension of the extracted feature vectors and \textit{n} is the number of visual networks as constructed for each fMRI session.   

\subsubsection*{Autoencoder network architecture}
Autoencoders belong to the category of feedforward neural networks extensively employed for reconstruction purposes. These networks compress the input data into a lower-dimensional latent space and then reconstruct the input from this compressed representation. The resulting latent space serves as a compact summary of the input, commonly referred to as the latent space representation. 
In general, an autoencoder comprises three layers: the encoder, code (also known as the bottleneck or latent space), and decoder. The encoder layer compresses the input data into a lower-dimensional representation in the latent space. Subsequently, the decoder layer reconstructs the encoded data back to its original dimension. Thus, the final output is derived from the latent space representation. Due to the smaller size of the input feature dimension, we utilize a 3-layer autoencoder with a latent space dimension of 4. This is shown in Figure~\ref{fig:deep-hybrid model}. For optimization, the mean squared error loss function is used to effectively reconstruct the input data from the latent space representation.

\subsubsection*{PCA-SVM model for classification}
In order to classify visual networks that represent COCO, ImageNet, and SUN across all sessions, we carry out the following steps. First, the obtained 4D latent space features of the autoencoder are directly fed into SVM with a higher-order polynomial kernel. The performance of SVM is further improved by applying PCA on 4D latent space features of the autoencoder. PCA is a widely used technique for dimensionality reduction, which learns a linear transformation to project the data into a new space. In this new space, the vectors of projections are defined by the variance of the data. By selecting a specific number of components that capture the majority of the variance in the dataset, PCA achieves effective dimensionality reduction. 
%This process allows to represent the data in a lower-dimensional space while preserving the most relevant information. 
PCA identifies and emphasizes the directions of maximum variance in the 4D latent space. Thus, applying PCA to the 4D latent space results in a new uncorrelated compact representation without loss of information. The proposed deep-hybrid architecture (Figure~\ref{fig:deep-hybrid model}) effectively combines the advantages of both the autoencoder and PCA techniques in order to build a refined, informative, and low-dimensional feature space. This increases the separation between different classes, making it easier for the SVM classifier to distinguish and classify the data accurately. \par
The SVM classification is conducted on the these PCA-reduced autoencoder latent space feature values, obtained with varying levels of explained variances. Explained variance quantifies how much of the variation in data is captured by each principal component. To visualize the relationship between the number of retained principal components and classification accuracy, we present a Scree-accuracy plot (Figure~\ref{fig:svm-all}) that helps to identify the optimal number of principal components needed to achieve the highest accuracy in the classification task. In the case of SVM classification on latent space PCA features, the RBF kernel is utilized. Given the limited number of the total visual networks, as mentioned before, binary classification with leave-one-out cross-validation is performed in order to categorize visual networks across the considered visual datasets.

%%%%%%%%%%%%%%%%%%%%%%%%%%%%%%%%%%%%%%%%%%%%%%%%%%%%%%%%%%%%%%%

\section*{Data availability}
The dataset analysed during the current study are available in BOLD5000 repository.
The dataset can be downloaded from this link: https://bold5000-dataset.github.io/website/download.html

\bibliography{Our_Paper}

\section*{Acknowledgement}
The authors would like to thank Mphasis F1 Foundation, cognitive computing grant to conduct research at International Institute of Information Technology (IIIT), Bangalore, Karnataka, India.

\section*{Author contributions statement}

All authors contributed to the study conception and design. D.B. conducted the experiment and analyzed the results under the guidance of N.S. at International Institute of Information Technology (IIIT), Bangalore. Problem statement of the research was formulated by D.B and N.S. The data preparation and interpretation of data for topological data analysis are performed by Y.R. and A.C. at IIIT Bangalore. The draft of the manuscript is written by D.B. and Y.R. and has been critically revised by N.S. and A.C. for important intellectual content. All authors reviewed and approved the final manuscript.

\section*{Competing interests}
The authors declare no competing interests.

\section*{Additional information}
Correspondence and requests for materials should be addressed to D.B.

%%%%%%%%%%%%%%%%%%%%%%%%%%%%%%%%%%%%%%%%%%%%%%%%%%%%%%%%%%%%%%%%%
%%%%%%%%%%%%%%%%%%%%%%%%%%%%%%%%%%%%%%%%%%%%%%%%%%%%%%%%%%%%%%%
%%%%%%%%%%%%%%%%%%%%%%%%%%%%%%%%%%%%%%%%%%%%%%%%%%%%%%%%%%%%%%

\end{document}